\documentclass[twocolumn]{aastex61}
\shorttitle{Helium Absorption in Escaping Exoplanet Atmospheres}
\shortauthors{Oklop\v ci\'c \& Hirata}

\begin{document}

\title{A New Window into Escaping Exoplanet Atmospheres:\\ 10830~\AA\ Line of Helium}

\correspondingauthor{Antonija Oklop\v ci\'c}
\email{antonija.oklopcic@cfa.harvard.edu}

\author[0000-0002-9584-6476]{Antonija Oklop\v ci\'c}
\affiliation{Institute for Theory and Computation, Harvard-Smithsonian Center for Astrophysics\\ 60 Garden Street, MS-51\\ Cambridge, Massachusetts 02138, USA}

\author{Christopher M. Hirata}
\affiliation{Center for Cosmology and Astroparticle Physics, Ohio State University\\ 191 West Woodruff Avenue\\ Columbus, Ohio 43210, USA}

\begin{abstract}

Observational evidence for escaping exoplanet atmospheres has been obtained for a few exoplanets to date. It comes from strong transit signals detected in the ultraviolet, most notably in the wings of the hydrogen Lyman-$\alpha$ (Ly$\alpha$) line. However, the core of the Ly$\alpha$ line is often heavily affected by interstellar absorption and geocoronal emission, limiting the information about the atmosphere that can be extracted from that part of the spectrum. Transit observations in atomic lines that are (a) sensitive enough to trace the rarefied gas in the planetary wind and (b) do not suffer from significant extinction by the interstellar medium could enable more detailed observations, and thus provide better constraints on theoretical models of escaping atmospheres. The absorption line of a metastable state of helium at 10830~\AA\ could satisfy both of these conditions for some exoplanets. We develop a simple 1D model of escaping planetary atmospheres containing hydrogen and helium. We use it to calculate the density profile of helium in the 2$^3$S metastable excited state and the expected in-transit absorption at 10830~\AA\ for two exoplanets known to have escaping atmospheres. Our results indicate that exoplanets similar to GJ~436b and HD~209458b should exhibit enhanced transit depths at 10830~\AA, with $\sim 8\%$ and $\sim 2\%$ excess absorption in the line core, respectively.

\end{abstract}



\section{Introduction} 
\label{sec:intro}

Close-in exoplanets give us a new insight into the mechanisms of atmospheric escape and mass loss. In highly irradiated planets, atmospheric escape can be very efficient and act collectively on the atmosphere as a fluid \citep[e.g.][]{OwenJackson2012}, instead of on a particle-by-particle basis. This hydrodynamic escape may be important for the planetary evolution, especially in low-mass planets that are more vulnerable to photoevaporation compared to massive planets with deep gravitational wells. This process has been proposed as an explanation for the observed paucity of short-period sub-Jupiter planets and the bimodal distribution of planet radii \citep{OwenWu2013, Lundkvist2016, Fulton2017}. Improving our knowledge of how hydrodynamic escape works and how it affects a broad range of atmospheres is therefore necessary for better understanding the demographics of planetary systems and their evolution.

Observational evidence for atmospheric escape has been obtained for a handful of exoplanets to date in the form of a strong absorption signal detected in the wings of the hydrogen Ly$\alpha$ line, but also in some UV lines of metals \citep{Vidal-Madjar2003,Vidal-Madjar2004, Lecavelier2010, Linsky2010,Fossati2010,Kulow2014}. The first observations of this kind were obtained for a transiting hot Jupiter HD~209458b by \citet{Vidal-Madjar2003}. Strong absorption in the wings of the Ly$\alpha$ line resulted in transit depth about an order of magnitude greater than the optical transit, suggesting that the observed cloud of hydrogen extends far away from the planet. Even greater transit depths in the wings of Ly$\alpha$ have been reported for a warm Neptune GJ~436b \citep{Kulow2014, Ehrenreich2015,Lavie2017}.

Several groups have developed theoretical models of escaping atmospheres \citep[e.g.][]{Lammer2003, Yelle2004, GarciaMunoz2007, Murray-Clay2009, Koskinen2010, Bourrier2013b, Tripathi2015, Salz2016, Carroll-Nellenback2017}. The methodology and the complexity varies greatly between these models, and hence their predictions, such as the expected mass loss rate, can differ by orders of magnitude. More detailed observations are required to place more stringent constraints on theoretical models. 

Ly$\alpha$ observations have been immensely valuable for providing evidence of atmospheric escape. However, there are inherent limitations of using this line. Ly$\alpha$ suffers from extinction by the interstellar medium and contamination from geocoronal emission, rendering the signal from the Ly$\alpha$ line core---and the valuable information content it might carry---irretrievable \citep[e.g.][]{Ehrenreich2015}.
    
Here, we investigate the possibility of probing the escaping atmospheres of exoplanets with the absorption line of helium at 10830~\AA. This line may provide a new wavelength window for studying the hydrodynamic escape and atmospheric mass loss. Its main advantages over the UV lines include weaker interstellar absorption\footnote{\citet{Indriolo2009} measured the column density of metastable helium through diffuse interstellar clouds and obtained an upper limit of $N\lesssim 10^9$~cm$^{-2}$, which is roughly three orders of magnitude lower than our prediction for escaping exoplanet atmospheres (see \autoref{fig:steady_state_profiles}, bottom panel).} and the possibility of ground-based observations.


\section{Helium Metastable State and the 10830~\AA\ Line}
\label{sec:metastable_helium}

The helium atom can exist in two configurations based on the relative orientation of its electrons' spin, singlet (anti-parallel) and triplet (parallel). The lowest-lying triplet level (2$^3$S) is almost decoupled from the singlet ground state (1$^1$S) because radiative transitions between them are strongly suppressed. Due to relativistic and finite-wavelength corrections to the magnetic dipole transition formula, the 2$^3$S triplet helium can radiatively decay into the singlet ground state with an exceptionally long lifetime of 2.2 hr \citep{Drake1971}. 

\begin{figure}
\centering
\includegraphics[width=0.42\textwidth]{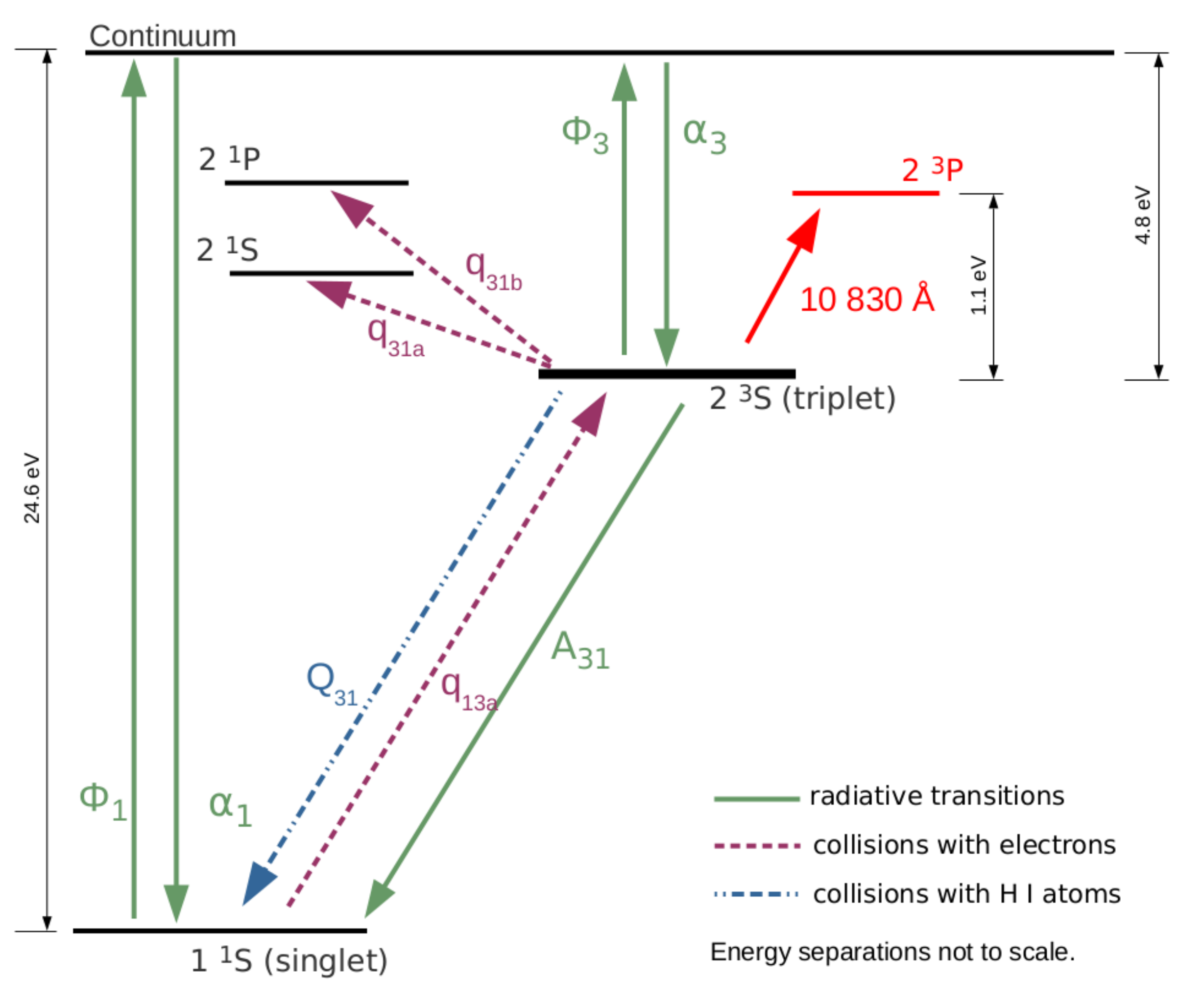}
\caption{Structure of the helium atom, indicating the radiative and collisional transitions included in our analysis. The transition shown in red depicts the 10830~\AA\ absorption line.}
\label{fig:atomic_structure}
\end{figure}

The 2$^3$S state can be populated by recombination\footnote{Around 75\% of helium recombinations result in the triplet configuration \citep{OsterbrockFerland}.} or by collisional excitation from the ground state (see \autoref{fig:atomic_structure}). Depopulation of this state progresses slowly making it metastable, and hence a promising origin of absorption lines. The resonance scattering from the metastable state to the $2^3P$ state produces an absorption line at 10830~\AA. This transition consists of three lines, two of which---at 10830.34~\AA\ and 10830.25~\AA---are practically indistinguishable, whereas the third component is at 10829.09~\AA.

The helium 10830~\AA\ line has been used to probe the dynamics of stellar winds \citep[e.g.][]{Dupree1992, Edwards2003} and outflows from active galactic nuclei \citep[e.g.][]{Leighly2011}, as well as the structure of the solar chromosphere and transition region \citep{Avrett1994,AndrettaJones1997,Mauas2005}. Even though this line has been identified as one of the most promising spectral signatures of exoplanet atmospheres \citep{SeagerSasselov2000}, no firm detections have been reported thus far \citep[see, e.g.,][]{Moutou2003}.

Most previous models of escaping atmospheres have focused on hydrogen or tracked helium only in terms of neutrals and ions, without modeling the metastable state. \citet{Turner2016} used \textsc{cloudy} to compile a list of $\sim60$ potentially interesting absorption lines for probing upper exoplanet atmospheres, including the 10830~\AA\ line.


\section{Methods and Results}
\label{sec:methods}

\subsection{Isothermal Parker Wind}

Our model is based on the assumption that upper layers of a hydrodynamically escaping atmosphere can be described by an isothermal Parker wind driven by gas pressure \citep{Parker1958,Watson1981}. Winds in which heating and cooling are due to radiative processes are close to being isothermal \citep{LamersCassinelli1999}. If the radiative heating/cooling processes operate on short timescales compared to the dynamical time of the system, the gas can self-regulate at a constant temperature.

A time-independent and spherically symmetric wind has a constant mass-loss rate determined by the equation of mass conservation:
\begin{equation}
\dot{M} = 4\pi r^2 \rho(r)v(r).
\label{eq:mass_cont}
\end{equation}
For an isothermal wind, the energy equation is
\begin{equation}
T(r) = T_0 \mbox{,}
\end{equation}
whereas the momentum equation is given by 
\begin{equation}
v\frac{dv}{dr} +\frac{1}{\rho}\frac{dp}{dr} + \frac{GM_{pl}}{r^2}=0 \mbox{.}
\end{equation}
We ignore the gravitational influence of the star and the Coriolis force. The momentum equation has a singularity at the so-called critical point, which in our case coincides with the sonic point, where the wind velocity equals the isothermal speed of sound
\begin{equation}
v_s = \sqrt{\frac{kT_0}{\mu m_H}},
\end{equation}
where $\mu m_H$ is the mean molecular weight. We assume that gas is made up of 90\% (by number) atomic hydrogen and 10\% helium\footnote{One of the planets whose escaping atmosphere we model (GJ~436b) has been suggested to have a helium-rich atmosphere \citep{Hu2015}. This could have important implications for 10830~\AA\ absorption; however, we leave the investigation of atmospheres with different compositions for future work.}. The radius of the sonic point is
\begin{equation}
r_s = \frac{GM_{pl}}{2v_s^2}.
\end{equation}

The velocity profile of the isothermal wind is given by \citep{LamersCassinelli1999}:
\begin{equation}
\frac{v(r)}{v_s}\exp{\left[-\frac{v^2(r)}{2v_s^2} \right]} = \left(\frac{r_s}{r} \right)^2\exp{\left(-\frac{2r_s}{r} +\frac{3}{2}\right)}.
\label{eq:parker_velocity}
\end{equation}
Using the mass continuity equation (\autoref{eq:mass_cont}), we get the equation for the density profile:
\begin{equation}
\frac{\rho(r)}{\rho_s} = \exp{\left[ \frac{2r_s}{r}-\frac{3}{2}-\frac{v^2(r)}{2v_s^2} \right]}.
\label{eq:parker_density}
\end{equation}

We use these expressions to set the density and velocity structure of the planetary wind. The free parameters in our model are the planet mass, radius, wind temperature and mass loss rate. We choose these parameters so that they match the properties of two well-studied exoplanets known to have escaping atmospheres: GJ~436b \citep{Butler2004}, a Neptune-sized planet orbiting an M-type star, and HD 209458b \citep{Henry2000, Charbonneau2000}, a hot Jupiter around a Sun-like star. Our model cannot predict the wind parameters such as the temperature and mass-loss rate, so we must assume their values. Our choice of $T_0 =5\times 10^3$~K and $\dot{M} = 2\times 10^{10}$~g~s$^{-1}$ for GJ~436b, and $T_0 =9\times 10^3$~K and $\dot{M} = 8\times 10^{10}$~g~s$^{-1}$ for HD~209458b, is guided by the results of \citet{Salz2016}, who used a hydrodynamics code coupled with \textsc{cloudy} \citep{Salz2015} to model atmospheric heating and wind launching for a number of known exoplanets. 

\begin{figure}
\centering
\includegraphics[width=0.5\textwidth]{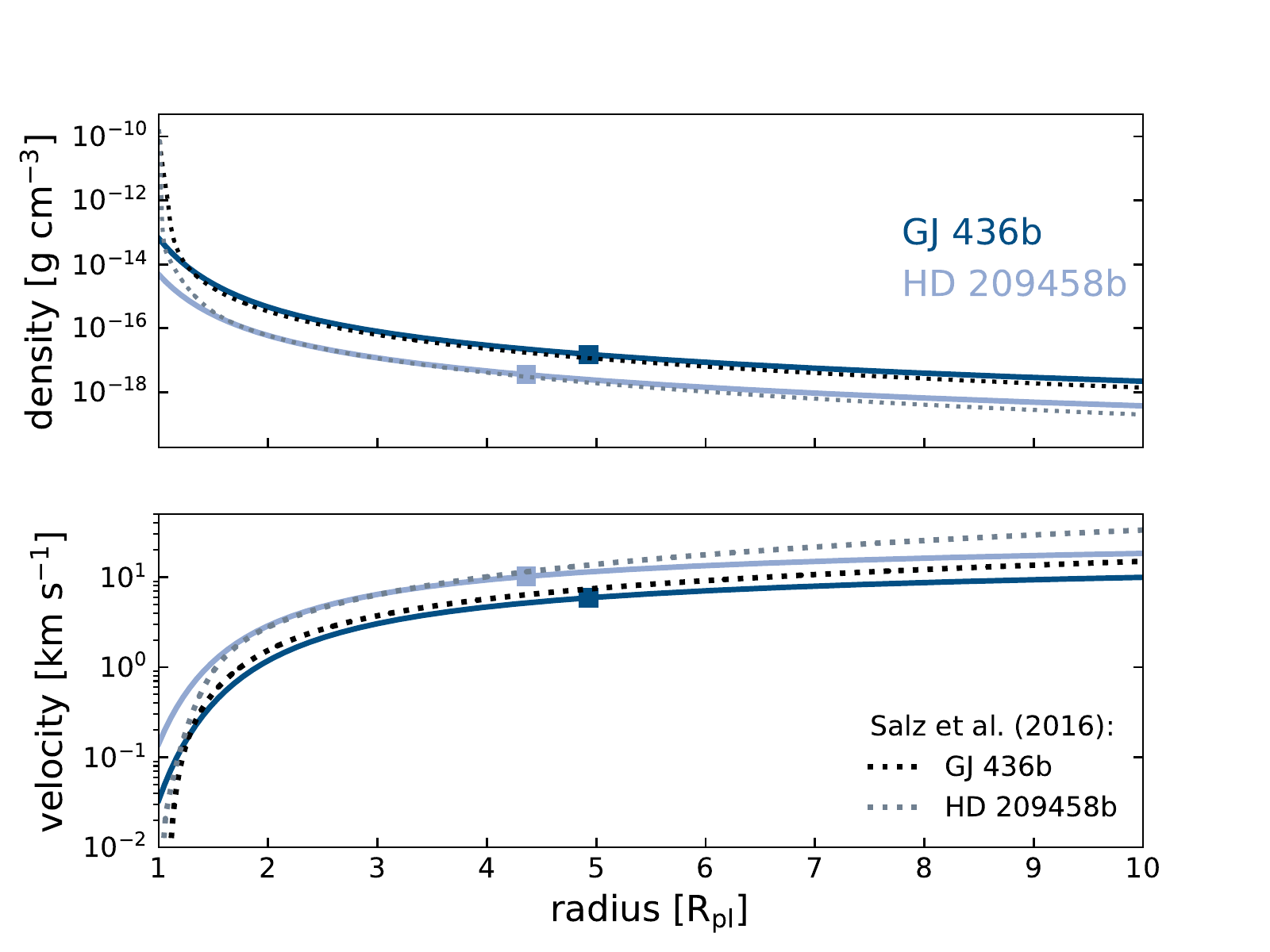}
\caption{Density and velocity profiles of the isothermal Parker wind for GJ~436b and HD~209458b. Squares mark the sonic point. Dotted lines show the corresponding profiles obtained by \citet{Salz2016}.}
\label{fig:parker_wind}
\end{figure}

In \autoref{fig:parker_wind}, we show our velocity and density profiles for both planets. Despite the simplicity of the isothermal model, the obtained atmospheric structure is very similar to the results of \citet{Salz2016} simulations, shown for comparison.

\subsection{Steady-state Hydrogen Distribution}

First, we calculate the radial density profile for hydrogen atoms and ions. The steady-state advection and recombination/ionization balance can be written as
\begin{equation}
\frac{1}{r^2}\frac{\partial}{\partial r}\left(r^2 n_{\text{\tiny H}^0} v\right) = n_{\text{\tiny H}^+}n_e \alpha_{rec} - n_{\text{\tiny H}^0}\Phi^\prime,
\label{eq:adv_ionH}
\end{equation}
where $n_{\text{\tiny H}^0}$, $n_{\text{\tiny H}^+}$, and $n_e$ denote the number density of neutral hydrogen, ionized hydrogen, and electrons, respectively, $v$ is the (radial) velocity of the outflow, $\alpha_{rec} =  2.59\times 10^{-13} (T_0/10^4)^{-0.7}$~cm$^3$~s$^{-1}$ is the case-B hydrogen recombination rate at $T_0$ \citep{OsterbrockFerland, Tripathi2015}, and $\Phi^\prime$ is the hydrogen photoionization rate calculated as
\begin{equation}
\Phi^\prime = \int_{\nu_0}^\infty \frac{F_{\nu}}{h\nu}a_\nu  e^{-\tau} d\nu \approx  e^{-\tau_0}\int_{\nu_0}^\infty \frac{F_{\nu}}{h\nu}a_\nu  d\nu \equiv e^{-\tau_0} \Phi
\label{eq:phi}
\end{equation}
where $\nu_0$ is the frequency corresponding to 13.6~eV photons. The hydrogen photoionization cross section $a_\nu$ is given by \citep{OsterbrockFerland}:
\begin{equation}
a_\nu = 6.3\times 10^{-18} \frac{\exp{\left(4-\frac{4\tan^{-1}\epsilon}{\epsilon}\right)}}{1-\exp{(-2\pi/\epsilon)}} \left(\frac{\nu_0}{\nu}\right)^4 \ \mbox{[cm$^2$]},
\end{equation}
where $\epsilon = \sqrt{\nu/\nu_0 -1}$. We calculate $\tau_0$ as 
\begin{equation}
\tau_0 = a_{0} \int_r^{\infty}n_{\text{\tiny H}^0} dr = \frac{0.9 a_{0}}{1.3 m_H}\int_r^\infty (1-f_\mathrm{ion})\rho(r) dr,
\label{eq:tau}
\end{equation}
where $\frac{0.9\rho}{1.3 m_{\text{\tiny H}}}$ is the number density of hydrogen nuclei, $f_\mathrm{ion}=n_{\text{\tiny H}^+}/(n_{\text{\tiny H}^0}+n_{\text{\tiny H}^+})$ is the ionized fraction of hydrogen (we assume $n_e=n_{\text{\tiny H}^+}$), and $a_0$ is the flux-averaged photoionization cross section given by
\begin{equation}
a_0 = \frac{\int_{\nu_{0}}^\infty F(\nu) a_\nu d\nu}{\int_{\nu_0}^\infty F(\nu) d\nu} ,
\label{eq:a0}
\end{equation}
where $F(\nu)$ is the stellar flux in units of photons/(cm$^2$ s Hz). Here we ignore the effects of radiative processes involving helium: optical depth in \autoref{eq:tau} does not account for helium absorption nor the hydrogen-ionizing photons produced by helium recombination.

$F_{\nu}$ in \autoref{eq:phi} denotes the stellar flux density. For GJ~436, we use the MUSCLES survey data, version 2.1 \citep{France2016,Youngblood2016,Loyd2016}. For HD~209458, a G0-type star, we use the SORCE solar spectral irradiance data from the LASP Interactive Solar Irradiance Datacenter\footnote{\url{http://lasp.colorado.edu/lisird/}}. Because HD~209458 is an inactive star \citep[and references therein]{Czesla2017}, we use the solar spectrum data recorded during a solar minimum\footnote{The population of the helium triplet state is very sensitive to the stellar UV flux. Using a solar spectrum from a more active period produces a stronger 10830~\AA\ absorption feature for HD~209458b.}. To fill in a gap in the data in the wavelength range $\sim 400-1150$\AA, we use the scaling relations between the Ly$\alpha$ flux and fluxes in EUV bands from \citet{Linsky2014}.

Using the continuity equation (\autoref{eq:mass_cont}), \autoref{eq:adv_ionH} can be written as
\begin{equation}
\frac{\partial f_\mathrm{ion}}{\partial r} = \frac{1-f_\mathrm{ion}}{v}\Phi e^{-\tau_0}  - \frac{0.9\rho }{1.3m_{\text{\tiny H}} v}f_\mathrm{ion}^2 \alpha_{rec}.
\end{equation}
The calculated fractions of neutral and ionized hydrogen as functions of altitude are shown in \autoref{fig:steady_state_profiles}.

\subsection{Steady-state Helium Distribution}
\label{sec:helium_distribution}

Similarly to our treatment of hydrogen, we can derive steady-state equations for the radial distribution of helium atoms:
\begin{eqnarray}
\nonumber v \frac{\partial f_1}{\partial r} &=& (1-f_1-f_3)n_e\alpha_1 + f_3 A_{31} - f_1\Phi_1e^{-\tau_1} \\
 &-& f_1n_e q_{13a} + f_3n_e q_{31a} + f_3 n_e q_{31b} + f_3 n_{\mathrm{H}^0}Q_{31},\\
\nonumber v  \frac{\partial f_3}{\partial r} &=& (1-f_1-f_3)n_e\alpha_3 - f_3 A_{31} - f_3\Phi_3e^{-\tau_3}+f_1n_e q_{13a} \\
&-& f_3n_e q_{31a}- f_3 n_e q_{31b} -f_3 n_{\mathrm{H}^0}Q_{31}.
\end{eqnarray}
$f_1$ and $f_3$ mark the fractions of helium in the (neutral) singlet and triplet state, respectively. $\alpha$ and $\Phi$ are the recombination and photoionization rate coefficients. We use the photoionization cross sections for the metastable triplet state from \citet{Norcross1971}. For the singlet state, we use the photoionization cross section from \citet{Brown1971}. Recombination rates are from \citet{OsterbrockFerland}. For simplicity, we assume that the number density of electrons is equal to the number density of ionized hydrogen atoms (i.e. we ignore the electrons produced by helium ionization, which can increase $n_e$ by up to $\sim 10\%$). Optical depths $\tau_1$ and $\tau_3$ are calculated using flux-averaged cross sections. For $\tau_1$, we take into account both helium and hydrogen absorption, and the threshold frequency in \autoref{eq:a0} now corresponds to 24.6 eV. For $\tau_3$, we integrate over frequencies between the 2$^3$S ionization threshold (4.8 eV) and the hydrogen-ionization threshold.

Transitions between singlet and triplet levels due to collisions with free electrons are described by coefficients
\begin{equation}
q_{ij} = 2.1\times 10^{-8} \sqrt{\frac{\mbox{13.6 eV}}{kT}}\exp{\left( -\frac{E_{ij}}{kT} \right)} \frac{\Upsilon_{ij}}{\omega_i} \ \ \mbox{[cm$^3$ s$^{-1}$],}
\end{equation}
where $\omega_i$ is the statistical weight of the initial level and $\Upsilon_{ij}$ is the effective collision strength from \citet{Bray2000}. The values obtained are $q_{13a}=4.5\times 10^{-20}$~cm$^3$ s$^{-1}$, $q_{31a}=2.6\times 10^{-8}$~cm$^3$ s$^{-1}$, and $q_{31b}=4.0\times 10^{-9}$~cm$^3$ s$^{-1}$. The metastable triplet level can also be depopulated by collisions with neutral hydrogen atoms via associative ionization
\begin{equation}
\mathrm{He} (2^3\mathrm{S}) + \mathrm{H} \rightarrow \mathrm{HeH}^+ + \mathrm{e}\ ,
\end{equation}
and Penning ionization
\begin{equation}
\mathrm{He} (2^3\mathrm{S}) + \mathrm{H} \rightarrow \mathrm{He} + \mathrm{H}^+ + \mathrm{e} \ \mbox{.}
\end{equation}
The combined rate coefficient for these processes is $Q_{31}\sim 5\times 10^{-10}$~cm$^{3}$~s$^{-1}$ \citep{RobergeDalgarno1982}. The radiative transition rate between the metastable and ground state is given by $A_{31}=1.272\times 10^{-4}$~s$^{-1}$ \citep{Drake1971}, which is one of the slowest rates in our reaction network.

\autoref{fig:atomic_structure} shows a schematic view of all the radiative and collisional transitions included in our calculation. The radiative transition between the metastable state and the 2$^3$P state---which is the origin of the 10830~\AA\ line---is not included in the reaction network because this process does not significantly depopulate the metastable state (i.e. it conserves the triplet configuration because an atom in the 2$^3$P state just decays back into the 2$^3$S state). We also neglect collisional ionization of helium, which is reasonable considering the assumed wind temperatures. We ignore direct transitions from the metastable to the ground state due to electron collisions because they are less probable than collisional transitions to the excited singlet states \citep{OsterbrockFerland}.

\begin{figure}
\centering
\includegraphics[width=0.45\textwidth]{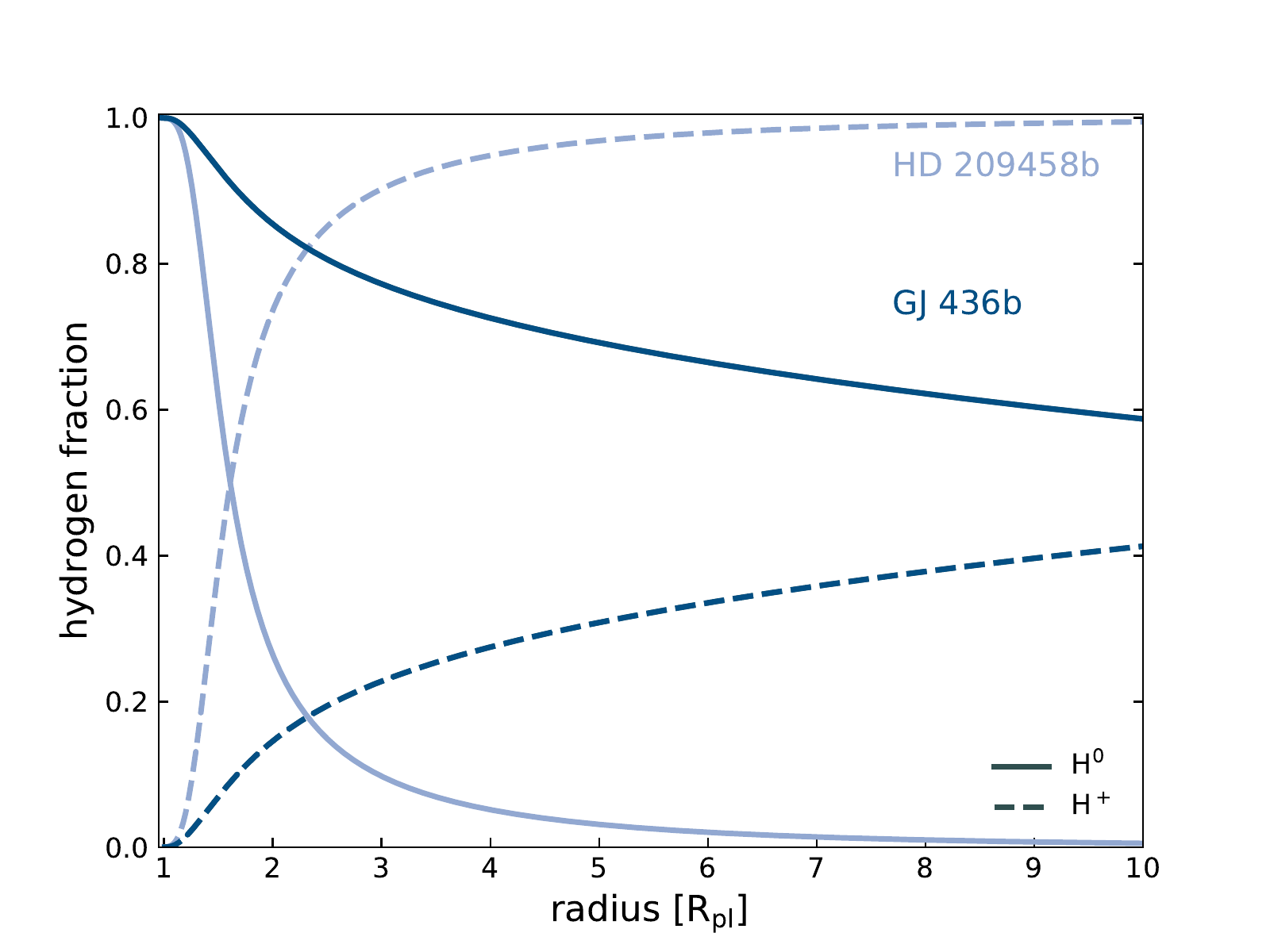}
\includegraphics[width=0.45\textwidth]{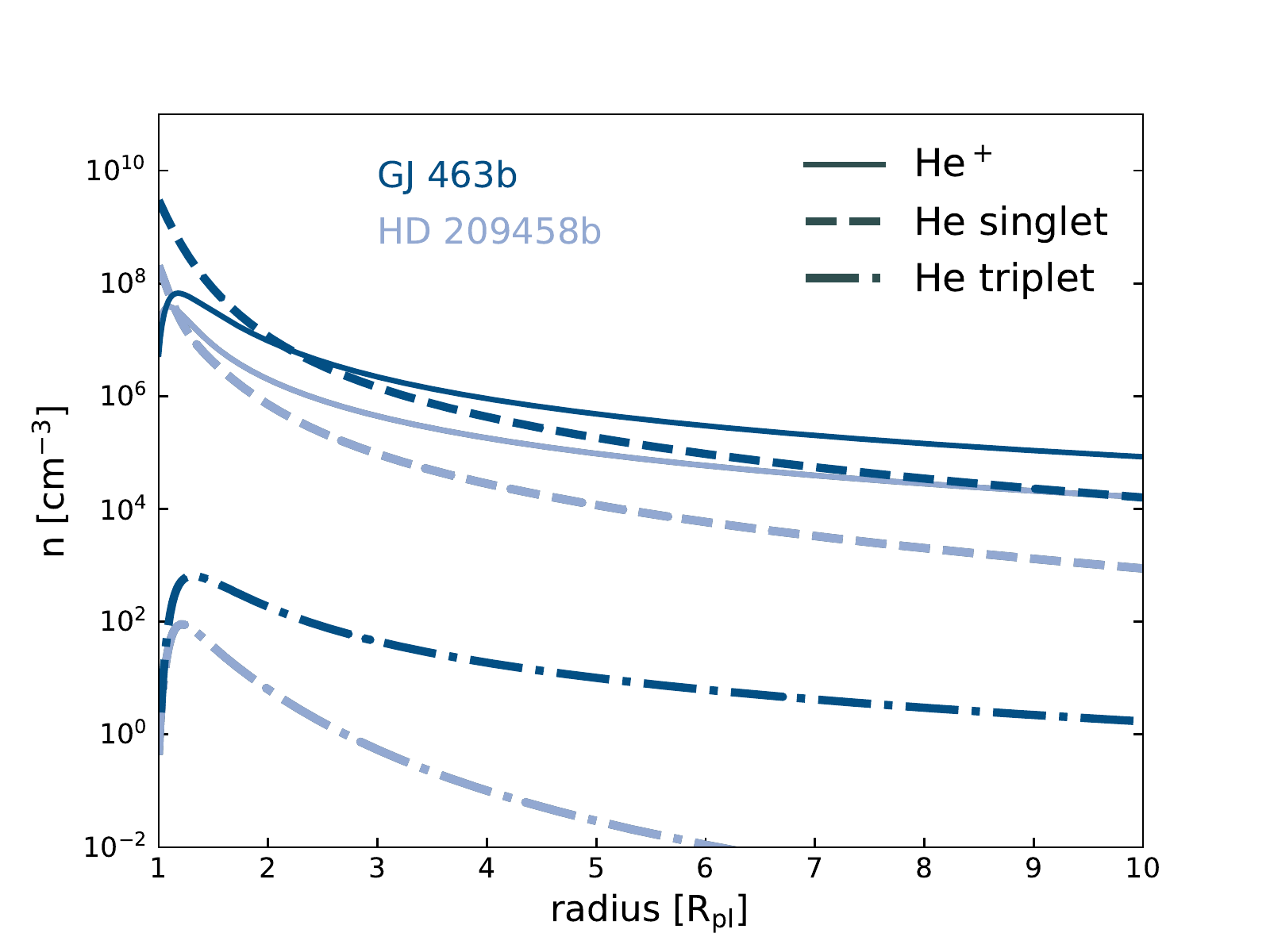}
\includegraphics[width=0.45\textwidth]{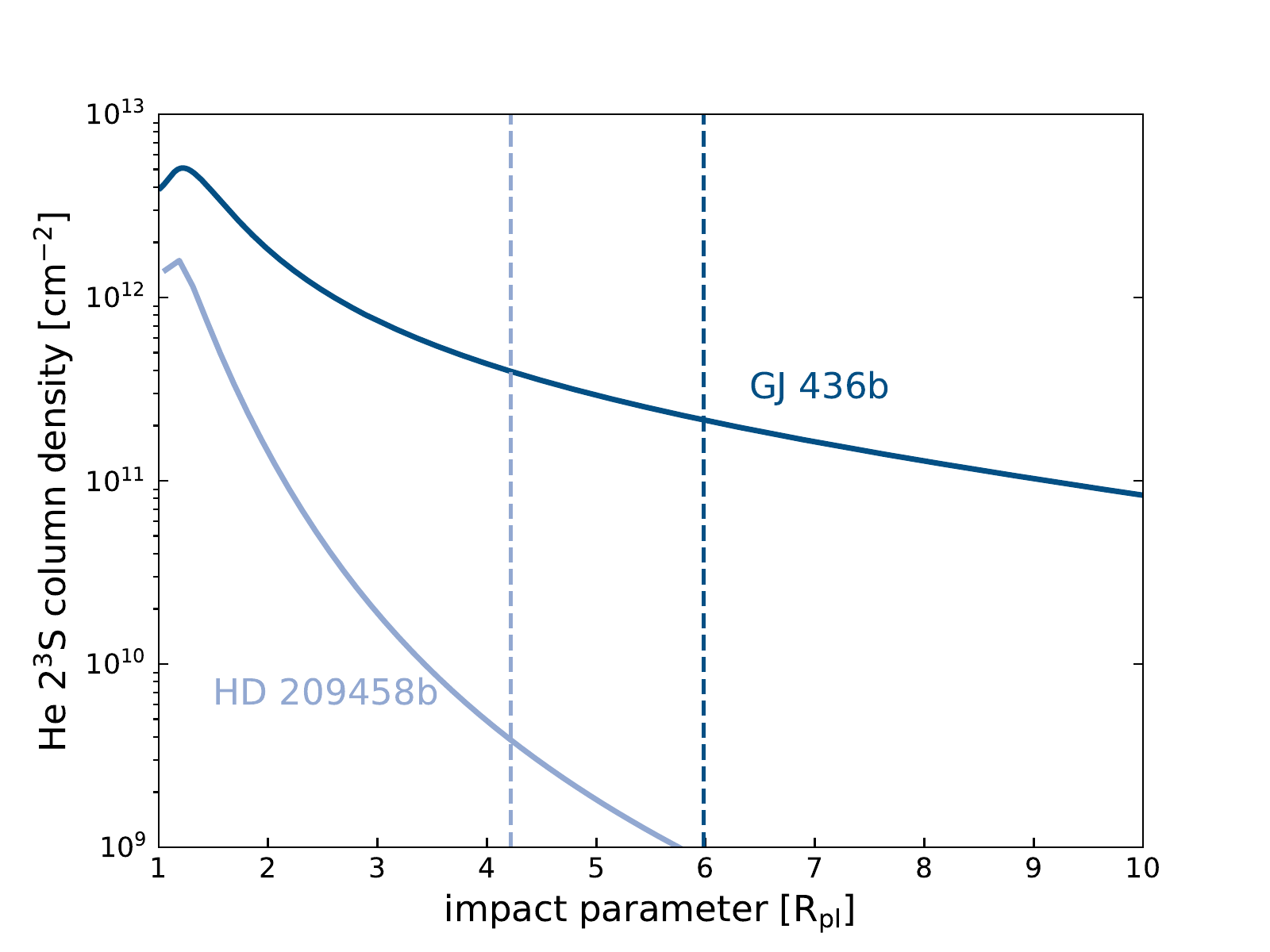}
\caption{Top: density profiles of neutral and ionized hydrogen. Middle: density profile of the ionized, ground state (singlet) and metastable (triplet) helium. Bottom: column density of metastable helium as a function of distance from the planet. Dashed lines mark the Roche-lobe radii.}
\label{fig:steady_state_profiles}
\end{figure}

The calculated density profiles of the singlet and triplet helium are shown in the middle panel of \autoref{fig:steady_state_profiles}. 

\subsection{Helium Line Absorption Profile in Transit}
\label{sec:line}

Finally, we calculate the expected absorption of stellar light at 10830~\AA\ due to the presence of an escaping planetary atmosphere. For absorption along a line of sight with an impact parameter $b$, the optical depth is given by \citep[see, e.g.][]{Koskinen2010}
\begin{equation}
\tau_\nu (b) = 2\int_b^{\infty} \frac{n_{3}(r)\sigma_0\Phi(\Delta \nu) r dr}{\sqrt{r^2 - b^2}},
\end{equation}
where $n_3$ is the number density of the metastable triplet helium. The absorption cross section is 
\begin{equation}
\sigma_0 = \frac{\pi e^2}{m_e c} f ,
\end{equation}
where $e$ and $m_e$ are the electron charge and mass, respectively, $c$ is the speed of light, and $f$ is the oscillator strength for the 10830~\AA\ transition (taken from the NIST Atomic Spectra Database\footnote{\url{https://www.nist.gov/pml/atomic-spectra-database}} for all three components of the line triplet). $\Phi (\Delta \nu)$ is the Voigt line profile\footnote{\url{https://scipython.com/book/chapter-8-scipy/examples/the-voigt-profile/} \citep{Hill2016}} with the half-width at half-maximum (HWHM) of the Gaussian part
\begin{equation}
\alpha = \sqrt{\frac{2\ln{2} k T_0}{m_{\mathrm{He}}}}\frac{\nu_0}{c},
\end{equation}
and the HWHM of the Lorentzian part $\gamma = A_{\text{\tiny{10830}}}/4\pi$, with the $A_{\text{\tiny{10830}}}= 1.0216\times 10^7$~s$^{-1}$ value from NIST.
The frequency offset from the line center $\Delta \nu = (\nu-\nu_0) - \frac{\nu_0}{c}v_{LOS}$ takes into account the line-of-sight component of the radial outflow velocity $v_{LOS}$ from our Parker wind velocity solution \citep[see, e.g.][]{Villarreal2014}. 

To compute the expected absorption line profile, we integrate $\tau_\nu (b)$ over the impact parameter from the planetary radius to the stellar radius, first by taking into account only the gas that is located within the planetary Roche lobe. We use the Roche-lobe height values from \citet{Salz2016}, 5.98~R$_\mathrm{pl}$ for GJ~436b and 4.22~R$_\mathrm{pl}$ for HD~209458b. The assumed spherical symmetry of the outflow is a reasonable approximation until the Roche radius. Beyond that, the wind can contribute to the absorption, but it may experience significant deviations from radial trajectories that can affect the line profile. To illustrate how much additional absorption could be caused by gas at larger radii, we also calculate the line profiles by taking into account gas at all radii.

\begin{figure}
\centering
\includegraphics[width=0.5\textwidth]{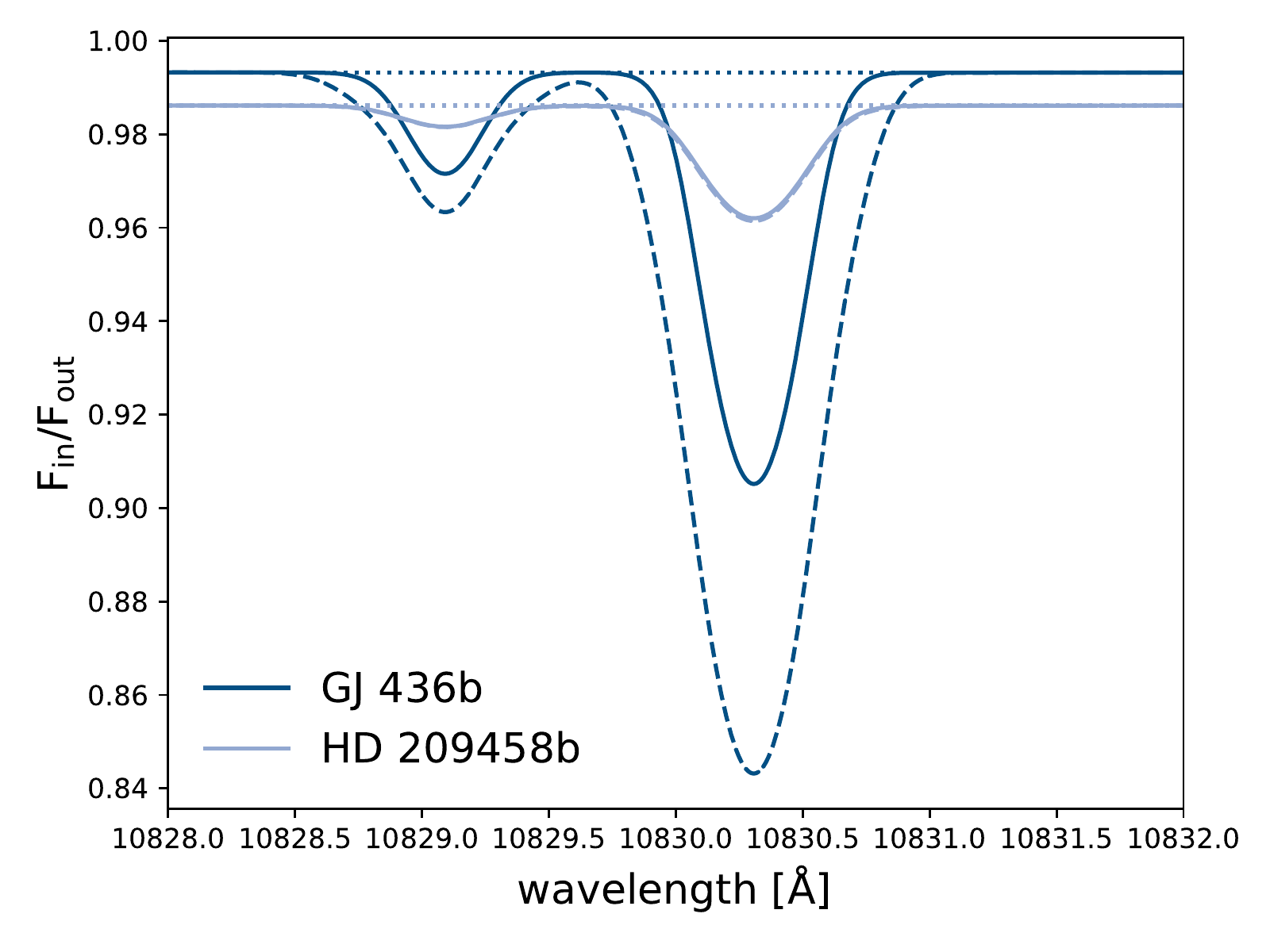}
\caption{Calculated transmission spectra for planets with properties similar to GJ 436b and HD 209458b, transiting across the center of the stellar disk. Solid lines show the absorption caused by the metastable helium within the planetary Roche lobe, whereas the dashed lines show the absorption caused by gas at all altitudes. Dotted lines mark the optical transit depths.}
\label{fig:transit_spectra}
\end{figure}

Our main result, shown in \autoref{fig:transit_spectra}, is the in-transit absorption calculated for planets with GJ~436b-like and HD~209458b-like properties, transiting across the center of their host star\footnote{GJ~436b does not transit across the center of its star, but is at a projected distance of $\sim0.85 R_*$. Consequently, a large fraction of the escaping gas does not transit the stellar disk and hence does not contribute to absorption.}. The equivalent width (EW) of the helium absorption feature (i.e. excess absorption in addition to the planet's optical transit depth) for GJ~436b is 0.047~\AA\ for gas up to the Roche radius (solid line) and 0.105~\AA\ for gas at all radii (dashed line). For HD~209458b, EW = 0.014~\AA. The dotted lines in \autoref{fig:transit_spectra} represent the optical transit depths, equal to $R_\mathrm{pl}^2/R_*^2 \approx 0.69\%$ and $1.4\%$, for GJ~436b and HD~209458b, respectively. 

We validate the procedure described in this section by calculating the in-transit absorption in the hydrogen Ly$\alpha$ line and comparing our results to observational and theoretical studies from the literature, as shown in \autoref{sec:lyalpha}.

\section{Discussion and Conclusions}
\label{sec:conclusions}

We develop a 1D model of escaping exoplanet atmospheres and use it to predict in-transit absorption signal in the 10830~\AA\ line of helium. If observed, the full absorption line profile at 10830~\AA\ would provide valuable information about the wind structure, including the base of the wind, which is currently lacking. This information would be complementary to that obtained in the extended wings of the Ly$\alpha$ line (because they probe different parts of the outflow), and would therefore help constrain theoretical models of escaping exoplanet atmospheres. 

\citet{SeagerSasselov2000} predicted a strong absorption at 10830~\AA\ for HD~209458b. Their study was focused on the stable, lower atmosphere of the planet; however, they state that extremely large transit signals could be expected from an extended exosphere. \citet{Moutou2003} measured the transit of HD~209458b at 10830~\AA\ and placed an upper limit of 0.5\% transit depth for a 3~\AA-wide line. Our model predicts a spectral feature of comparable EW, but much deeper and narrower. Given the significant differences in the line morphology and the fact that EWs differ by $\lesssim$6\%, a detailed analysis would be required to determine whether our prediction is consistent with \citet{Moutou2003} observation. However, this level of discrepancy could also be caused by the uncertainties in the input parameters we assume. \citet{Turner2016} model the upper exoplanet atmosphere as a static slab of uniform-density ($n\sim 10^9$~cm$^{-3}$) gas. They predict transit depths for dozens of spectral lines, including the 10830~\AA\ line, for which they obtain a transit depth of $\sim$0.3\% over a $\sim 50$~\AA-wide bin (J. Turner, priv. comm.). That is about an order of magnitude higher EW than our result for HD~209458b, which could be explained by differences in model geometry. They assume that high-density gas covers $\sim$40\% of the stellar disk in projection, whereas our density profile drops below $10^9$~cm$^{-3}$ at $\sim$1.5~R$_\mathrm{pl}$, occulting only $\sim 3$\% of the stellar disk.

Our results indicate that exoplanets like GJ~436b and HD~209458b should show enhanced absorption at 10830~\AA\ due to the presence of helium in the metastable state in their planetary wind. Based on our calculations, a GJ~436b-like planet is a more promising candidate for detecting the 10830~\AA\ line than a planet like HD~209458b. This is caused by the combined effect of (a) the fact that escaping atmospheres in planets with low gravitational potential tend to be denser at high altitudes, (b) different levels of helium ionization, (c) more favorable flux ratio of radiation responsible for populating versus depopulating the metastable state, and (d) differences in the hydrogen-ionizing radiation, which controls the hydrogen ionization fraction and electron density and, consequently, the relative contribution of different collisional depopulation mechanisms of the metastable state. We leave a more detailed investigation of how various stellar and planetary properties affect the expected absorption at 10830~\AA\ for future work.

In deriving our wind model, we make a number of simplifying assumptions. We perform our calculation in 1D and assume the outflow is spherically symmetric. According to the results of \citet{Khodachenko2015}, this approximation is valid at altitudes up to a few planetary radii, where most of our signal comes from. 3D simulations find dayside to nightside differences in wind properties, and a `cometary' tail of wind material at large distances behind the planet \citep[e.g.][]{Tripathi2015, Christie2016, Schneiter2016, Carroll-Nellenback2017}, which our 1D model cannot reproduce. We do not model the interaction between the escaping material and the stellar wind \citep[e.g.][]{Bourrier2016}. Simulations of atmospheric heating and wind launching suggest that planetary winds are not strictly isothermal, which is another assumption that we make. Although this assumption does not greatly affect the wind density and velocity structure (\autoref{fig:parker_wind}), it might have more subtle effects on the spectral line profile. A major limitation of our model is that we have to assume values for the wind temperature and mass-loss rate, and cannot predict them in a self-consistent way. 

The main advantage of our model is that it is computationally less expensive than hydrodynamic simulations. This will allow us to explore a wide range of planetary parameters and stellar spectral types in our future work, in order to identify what part of the parameter space is most promising for producing strong absorption signals in the 10830~\AA\ line. Once we identify the best candidates, more computational resources can be invested into performing detailed studies of these systems, using 2D or 3D simulations with more physically motivated treatment of wind launching.

\acknowledgments
We thank the anonymous referee for providing very helpful comments. We thank David Charbonneau, Andrea Dupree, and Jessica Spake for insightful conversations. AO acknowledges support from an ITC Fellowship. CMH is supported by NASA, NSF, the U.S. Department of Energy, the David \& Lucile Packard Foundation, and the Simons Foundation.

\software{matplotlib \citep{Hunter2007}, numpy \citep{VanDerWalt2011}, scipy \citep{Jones2001}}

\appendix
\section{Ly$\alpha$ Absorption}
\label{sec:lyalpha}

In addition to the helium 10830~\AA\ line, we can use our model to calculate the absorption in the hydrogen Ly$\alpha$ line. Following the procedure described in \autoref{sec:line}, we calculate the optical depth to neutral hydrogen (protium and deuterium, assuming a deuterium fraction of $2.25\times 10^{-5}$, as measured in Jupiter by \citet{Lellouch2001}). We calculate the hydrogen Ly$\alpha$ line profiles using the wavelengths, oscillator strengths, and natural broadening parameters from the NIST database.

\begin{figure*}
\centering
\includegraphics[width=0.48\textwidth]{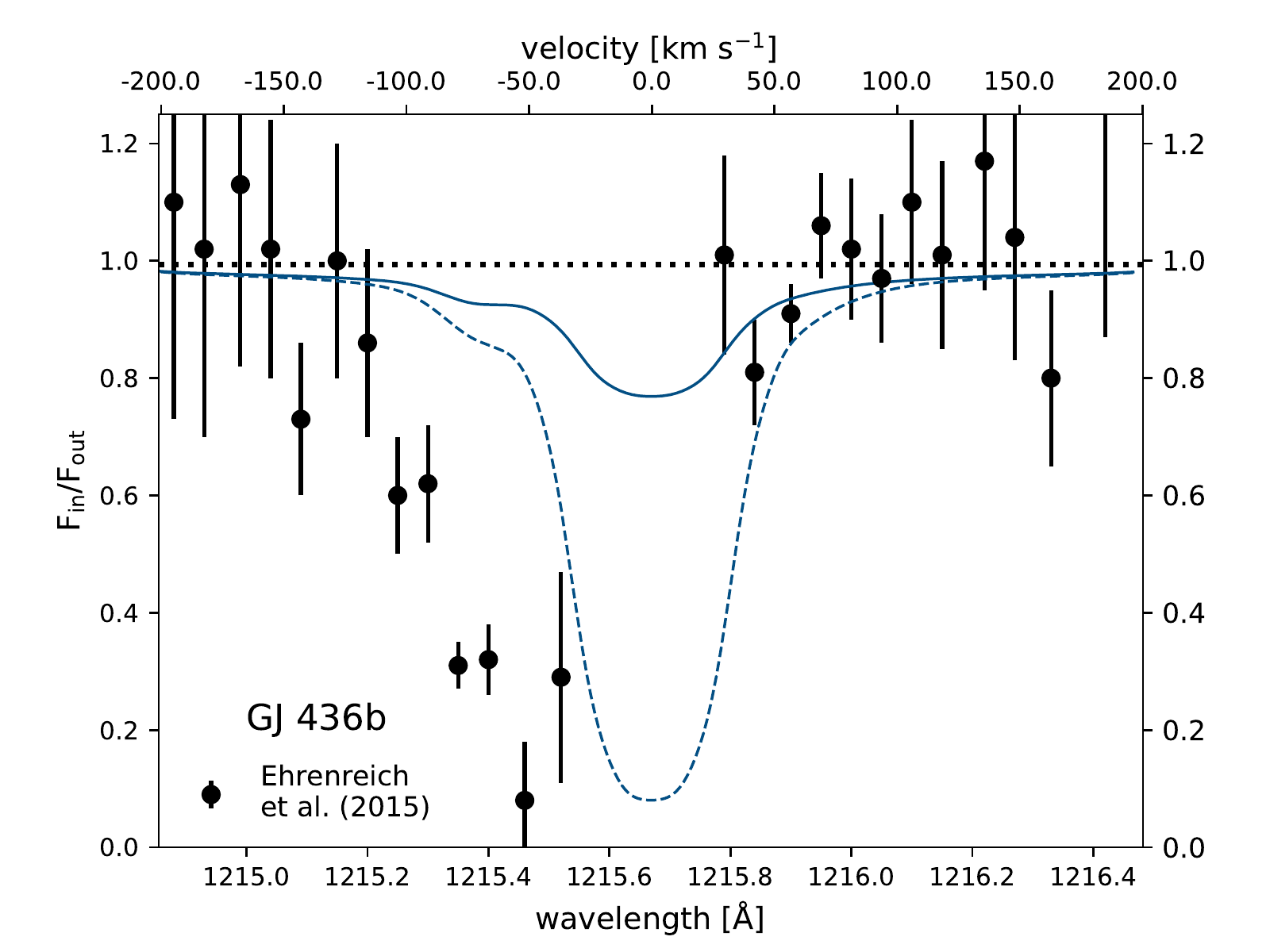}
\includegraphics[width=0.48\textwidth]{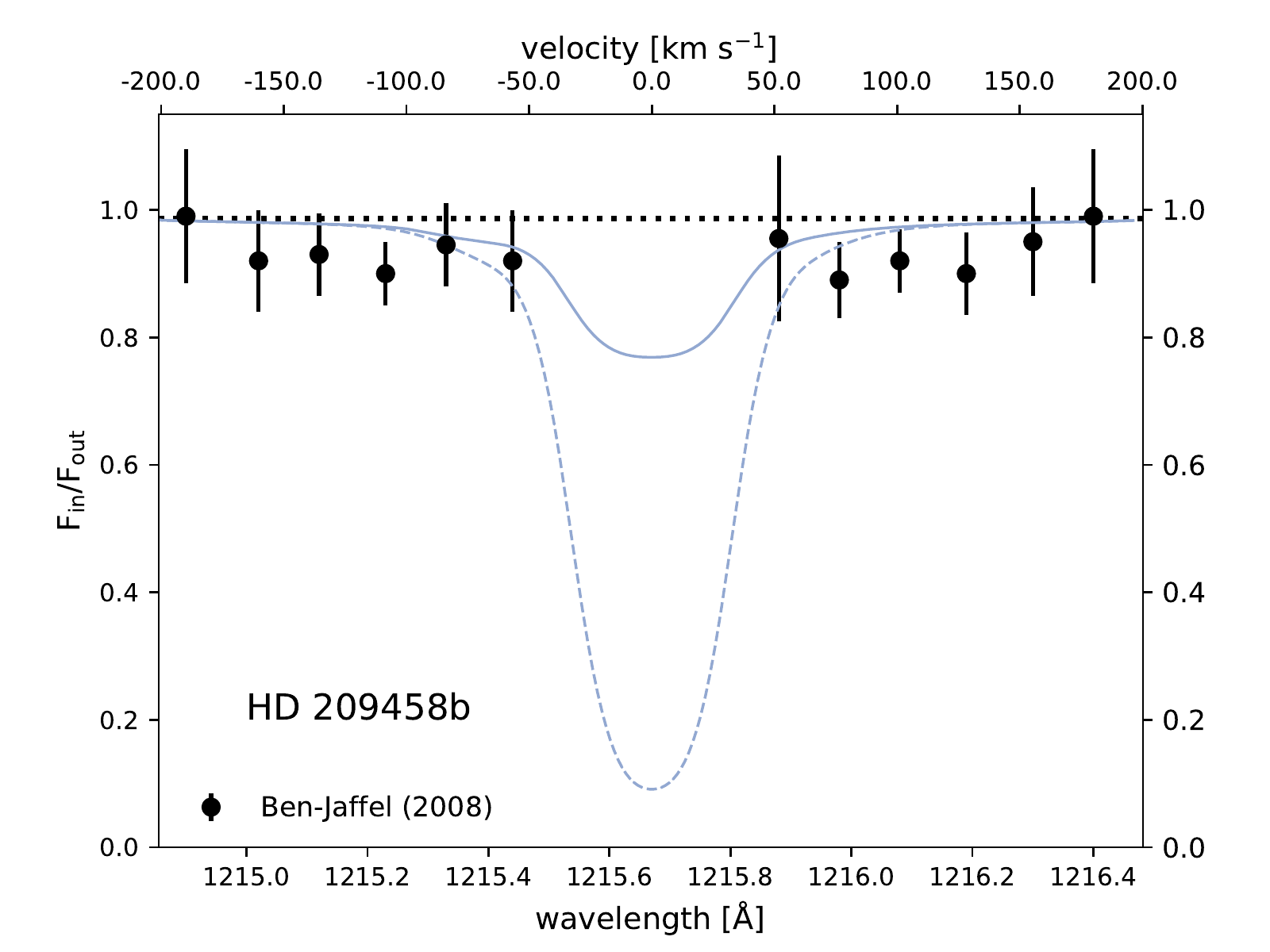}
\caption{Ly$\alpha$ transmission spectra for planets with properties like GJ 436b and HD 209458b, transiting across the center of the stellar disk. Solid lines show the absorption caused by hydrogen located within the Roche radius, whereas the dashed lines take into account gas at all altitudes. Transit observations are shown for comparison. The spectral region near the line core that is affected by the interstellar absorption and geocoronal emission is omitted from the observational data.}
\label{fig:lyalpha}
\end{figure*}

In \autoref{fig:lyalpha} we show the obtained line profiles for both planets (convolved with the HST/STIS line-spread function for G140M grating and aperture of 52\arcsec$\times$0\arcsec.1) and compare them with observations from \citet{Ben-Jaffel2008} and \citet{Ehrenreich2015}. Our simple wind model cannot fully explain the observed absorption in the \textit{wings} of the Ly$\alpha$ line due to the simplifications discussed in \autoref{sec:conclusions}. The Ly$\alpha$ line \textit{center}---where our model should be more reliable---is observationally unattainable due to the interstellar absorption and geocoronal emission. Our results agree reasonably well with the results of theoretical studies by \citet{Ben-Jaffel2010} and \citet{Salz2016}.

\end{document}